\documentclass[twocolumn]{aastex63}
\usepackage{graphicx,subfigure,enumerate,pstricks,latexsym,longtable,amsfonts,amsmath}
\usepackage{soul}
\usepackage{lipsum}

\newcommand{\gama}{$\gamma$}

\newcommand{\beq}{\begin{equation}}
\newcommand{\eeq}{\end{equation}}

\shorttitle{blazar chaos}
\shortauthors{Bhatta \& P{\'a}nis \& Stuchl{\'i}k}

\begin{document}
\title{Deterministic Aspect of the $\gamma$-ray Variability in Blazars  }

\correspondingauthor{Gopal Bhatta}
\email{gopal@oa.uj.edu.pl}

\author{Gopal Bhatta}
\affiliation{Astronomical Observatory of the Jagiellonian University \\
ul. Orla 171 \\
 30-244 Krak\'ow, Poland}

\author{Radim P{\'a}nis}
\affiliation{Research Centre for Theoretical Physics and Astrophysics, Institute of Physics\\
Silesian University in Opava\\
 Bezru{\v c}ovo n{\'a}m.13, CZ-74601 Opava, Czech Republic}

\author{Zden\v{e}k Stuchl{\'i}k}
\affiliation{Research Centre for Theoretical Physics and Astrophysics, Institute of Physics\\
Silesian University in Opava\\
Bezru{\v c}ovo n{\'a}m.13, CZ-74601 Opava, Czech Republic}

\begin{abstract}
Linear time series analysis, mainly  the Fourier transform based methods, has been quite successful in extracting information contained in the ever-modulating light curves (Lcs) of active galactic nuclei, and thereby contribute in characterizing the general features of supermassive black hole systems.  In particular, the statistical properties of $\gamma$-ray variability of blazars are found to be fairly represented by flicker noise in the temporal frequency domain. However, these conventional methods have not been able to fully encapsulate the richness and the complexity displayed in the light curves of the sources.
In this work, to complement our previous study on the similar topic, we perform non-linear time series analysis of the decade-long Fermi/LAT observations of 20 $\gamma$-ray bright blazars.  The study  is motivated to address one of the most relevant queries that whether  the dominant dynamical processes leading to the observed $\gamma$-ray variability are of deterministic or stochastic nature. For the purpose, we perform Recurrence Quantification Analysis of the blazars and directly measure the quantities which suggest that the dynamical processes in blazar could be  a combination of deterministic and stochastic processes, while some of the source light curves revealed significant deterministic content. The result with possible implication of strong disk-jet connection in blazars could prove to be significantly useful in constructing models that can explain the rich and complex multi-wavelength observational features in active galactic nuclei.  In addition, we estimate the dynamical timescales, so called  ``trapping timescales'', in the order of a few weeks.

\end{abstract}

\keywords{accretion, accretion disks --- radiation mechanisms: non-thermal, \gama-ray --- galaxies: active --- galaxies: jets, method: non-linear time series analysis}


\section{Introduction} \label{sec:intro}

Blazars are extra-galactic, supermassive black hole systems that display relativistic jet closely pointed towards the Earth.  The sources come mainly in two flavors:  flat-spectrum radio quasars (FSRQ), the more luminous  kind that shows emission lines over the continuum, and BL Lacertae (BL Lac) sources, the less powerful  objects which show weak or no such lines. The current and widely accepted models paint a spectacular picture of the blazar systems: as plasma material swirls inward close to the supermassive black holes of the masses in the order $\sim 10^9\rm M_\odot$, the magnetic field in conjunction with the fast rotation of the supermassive black hole contributes to the launching of the bi-polar relativistic jets  which then travel up to Mpc scale distance \citep{Blandford2019,Blandford1977}. While the jets plough through the intergalactic medium, any small velocity gradient can lead to formation of shock waves and consequently create favorable condition for the  violent episodes giving rise to the large amplitude flares as observed in the light curves \citep[see][]{2016Galax...4...37M}. It is believed that the jet contents could be dominated by the Poynting flux such that  the relativistic  electrons give rise to synchrotron emission; and  the accelerated charged particles upscatter either the population of co-spatial  synchrotorn photons (Synchrotron Self-Compton model; e.g., \citealt{Maraschi1992,Mastichiadis2002}) or the lower energy photons coming from various parts e.g.  accretion disk (AD; \citealt{Dermer1993}), broad-line region (BLR; \citealt{Sikora1994}), and dusty torus (DT; \citealt{Blazejowski2000}) -- External Compton model. As the result, blazars become dominant sources of high energy emission along with possible extra-galactic sources of neutrinos \citep[see][]{2018Sci...361.1378I,2018Sci...361..147I}.

Flux variability in diverse temporal and spatial frequencies is one of the defining and fascinating properties of blazars (e. g. \gama-ray; \citealt{Rajput2020}, X-ray; \citealt{Bhatta2018c}, optical; \citealt{Bhatta2018a}).  The power spectral density analysis reveals that the statistical nature of the blazar \gama-ray variability can be   well  described by power-law  type noise with mostly single power-law index \citep[][and the references therein]{Bhatta2020}. A number of models attempt to explain the variability linking its origin to various mechanisms, e.g.,  magnetohydrodynamic instabilities in the jets \citep[e.g.][]{bhatta13, Marscher14}, shocks traveling down the turbulent jets \citep[e.g.][]{Marscher1985,Bottcher2010}, magnetic reconnection in the turbulent jets \citep{Sironi2015,Werner2016}, and relativistic effects due to jet orientation\citep[e.g.][]{Camenzind92,Raiteri2017}. However, working out the exact details of the underlying processes has been part of ongoing research. 

In general, the time series analysis, mostly Power Spectrum Density (PSD) analysis, are treated as one of the most powerful tools in characterizing the statistical nature of the observed variability. However, usage of such analyses are limited to the second-order moments of the flux distribution and static properties of  the light curves. Consequently, the methods fail to incorporate the information about the inherent non-linearity and non-stationarity which is contained in the higher order moments and which directly reflect into the dynamical nature of the black hole systems \citep[see][]{Shoji2020,Zbilut2008,Green1999}.  Moreover, the attempts to constrain the observed variability in the blazar within the framework of linear
stochastic systems probe into the randomly occurring flaring episodes, such as  local fluctuations in the viscosity, accretion rate at  the accretion disc, and/or stochastic shock events prevailing the jet regions. Such linear stochastic  changes are not likely to affect global  perturbations which ultimately materialize in the observed flux changes in the sources.
 On the other hand,  the observational feature such as RMS-flux relation and log-normal flux distribution \citep[e. g.][]{Bhattacharyya2020,Bhatta2020,Uttley2005} point out to the non-linear dynamics  inherent in the disk-jet systems, and therefore explore into  the processes that lead to global perturbations giving rise to the instabilities that persist and remain coherent over the entire system \citep[however, for a shot noise interpretation of such observations see][]{Scargle2020}. Studies of black hole systems taking the non-linear time series approach to the AGN light curves can be found in several works \citep[e.g.][]{Shoji2020,Bachev2015,Leighly1997,2020arXiv200102800P}. Besides, the non-linear time series analysis can be used to distinguish sources which have similar set of the non-linear properties as well as measure characteristic timescales, e. g. trapping timescales which represents an average time a system spends on a particular state \citep[see][]{Marwan2002}.

More importantly, the query whether the basic nature of the variability should be treated as stochastic or deterministic stands out as one of the most relevant questions to be asked (see \citealt{Kiehlmann2016}; and in the context of microquasars see \citealt{2016A&A...591A..77S}). The answer to such queries has far-reaching impact in our attempts to constrain that physical process that lead the multi-timescale variability, e. g. the physical conditions prevailing the innermost regions of blazar jets, the nature of the dominant particle acceleration and energy dissipation mechanism, magnetic field geometry, jet content, etc. It is most likely that the roots of variability phenomenon can be related to the non-linear magnetohydrodynamical flows at the accretion-jet systems that are governed by the combined effect of the ambient magnetic field and the rotation of the innermost regions around the supermassive black holes. In such scenario, non-linear time series analysis estimating the changes in the dynamical states of the system can contribute in establishing a strong connection between the accretion disk and  the jet in radio-loud AGN systems \citep[see][for observational signature of the disk-jet connection]{Bhatta2018b}.

In this work, we carry out non-linear time series analysis of 20 blazars utilizing decade long Fermi/LAT light curves presented in our previous work \citep[see][]{Bhatta2020}. In Section \ref{sec:3}, the details of the analysis, in particular, Recurrence Quantification Analysis (RQA), which provides various measures including determinism, predictability, and entropy, is discussed in detail. In addition, the results of the analyses on the \gama-ray light curves are also presented. Then discussion on the results along with their possible implications on the nature of \gama-ray emission from the sources are presented in Section \ref{sec:4}, and  finally the  conclusions of the study summarized are in Section \ref{sec:5}.

\section{Source Sample \label{sec:2}}

The source sample consists of 20 \gama-ray bright blazars such that weekly binned light curves can be constructed\footnote{The Fermi/LAT data acquisition and processing are discussed in \citet{Bhatta2020}}. The sources are listed in Column 1 of Table \ref{table:1} along with their positions in the sky, right ascension (Col. 2) and declination (Col. 3), 3FGL catalog names (Col. 4), source class ( Col. 5), and their red-shits as listed in NED\footnote{\url{https://ned.ipac.caltech.edu/}.}

\begin{deluxetable*}{llllll}
\tablecaption{The source sample of the 
blazars included in the study \label{table:1}}
\tablewidth{500pt}
\tabletypesize{\scriptsize}
\tablehead{
\colhead{Source name} & \colhead{R.A. (J2000)} & 
\colhead{Dec. (J2000)} & \colhead{3FGL name} & 
\colhead{Source class} & \colhead{Red-shift} \\
} 
\colnumbers
\startdata
	3C 66A 	&	 $02^h22^m41.6^s$ 	&	 $+43^d02^m35.5^s$ 	&	3FGL J0222.6+4301 	&	BL Lac 	&	0.444	\\
	AO 0235+164 	&	 $02^h 38^m38.9^s$ 	&	 $+16^d 36^m 59^s$ 	&	3FGL J0238.6+1636	&	BL Lac 	&	0.94	\\
	PKS 0454-234 	&	 $04^h 57^m03.2^s$ 	&	 $-23^d 24^m 52^s$ 	&	3FGLJ0457.0-2324	&	BL Lac 	&	1.003	\\
	S5 0716+714 	&	$07^h21^m53.4^s$ 	&	 $+71^d20^m36^s$ 	&	3FGL J0721.9+7120 	&	BL Lac 	&	0.3	\\
	Mrk 421 	&	 $11^h04^m273^s$ 	&	 $+38^d12^m32^s$ 	&	3FGLJ1104.4+3812 	&	BL Lac 	&	0.03	\\
	 TON 0599 	&	$11^h59^m31.8^s$ 	&	 $+29^d14^m44^s$ 	&	3FGL J1159.5+2914 	&	BL Lac 	&	0.7247	\\
	ON +325 	&	 $12^h17^m52.1^s$ 	&	 $+30^d07^m01^s$ 	&	3FGL J1217.8+3007 	&	BL Lac 	&	0.131	\\
	W Comae 	&	 $12^h 21^m31.7^s$ 	&	 $+28^d 13^m 59^s$ 	&	3FGL J1221.4+2814	&	BL Lac 	&	0.102	\\
	4C +21.35 	&	 $12^h24^m54.4^s$ 	&	 $+21^d22^m46^s$ 	&	3FGLJ1224.9+2122 	&	FSRQ 	&	0.432	\\
	3C 273 	&	 $12^h29^m06.6997^s$ 	&	 $+02^d03^m08.598^s$ 	&	3FGL J1229.1+0202 	&	FSRQ 	&	0.158	\\
	3C 279 	&	 $12^h56^m11.1665^s$ 	&	 $-05^d47^m21.523^s$ 	&	3FGL J1256.1-0547 	&	FSRQ 	&	0.536	\\
	PKS 1424-418 	&	 $14^h27^m56.3^s$ 	&	 $-42^d06^m19^s$ 	&	3FGLJ1427.9-4206 	&	 FSRQ	&	1.522	\\
	PKS 1502+106 	&	 $15^h 04^m25^s.0$ 	&	 $+10^d 29^m 39^s$ 	&	3FGLJ1504.4+1029	&	FSRQ 	&	1.84	\\
	4C+38.41 	&	 $16^h35^m15.5^s$ 	&	 $+38^d08^m04^s$ 	&	3FGL J1635.2+3809 	&	FSRQ 	&	1.813	\\
	Mrk 501 	&	 $16^h53^m52.2167^s$ 	&	 $+39^d45^m36.609^s$ 	&	3FGL J1653.9+3945 	&	 BL Lac 	&	0.0334	\\
	1ES 1959+65 	&	 $19^h59^m59.8521^s$ 	&	 $+65^d08^m54.652^s$ 	&	3FGL J2000.0+6509 	&	BL Lac 	&	0.048	\\
	PKS 2155-304 	&	 $21^h58^m52.0651^s$ 	&	 $-30^d13^m32.118^s$ 	&	3FGL J2158.8-3013 	&	BL Lac 	&	0.116	\\
	BL Lac 	&	$22^h02^m43.3^s$ 	&	 $+42^d16^m40^s$ 	&	3FGL J2202.7+4217 	&	BL Lac 	&	0.068	\\
	CTA 102 	&	 $22^h32^m36.4^s$ 	&	 $+11^d43^m51^s$ 	&	3FGL J2232.5+1143 	&	FSRQ 	&	1.037	\\
	3C 454.3  	&	 $22^h53^m57.7^s$ 	&	 $+16^d08^m54^s$ 	&	3FGL J2254.0+1608 	&	FSRQ 	&	0.859	\\
\enddata
\end{deluxetable*}

\section{Analysis \label{sec:3}}
In order to further explore the nature of variability in \gama-ray light curves of the sample blazars, we adopted a number of approaches to the chaos study. The description of the methods and the corresponding results of the analyses are presented below.

\begin{table*}[ht]
\centering
\begin{tabular}{llrrrrrrrrrr}
  \hline
 & source & Len & $\tau$ & m & mD & mL & mEN & msD & msL & msEN \\ 
  \hline
1 & W Comae & 208 & 2 & 8 & 0.6507 & 118.0000 & 0.0000 & 20.0 & 12.5 & 1.0  \\ 
  2 & 3C 454.3 & 462 & 9 & 9 & 0.5169 & 7.8129 & 1.2666 & 19.0 & 1.5 & 17.5  \\ 
  3 & AO 0235+164 & 273 & 4 & 7 & 0.4178 & 183.0000 & 0.0000 & 16.0 & 13.5 & 1.0  \\ 
  4 & 4C+21.35 & 373 & 3 & 9 & 0.4086 & 12.8359 & 1.2823 & 17.0 & 5.5 & 16.5  \\ 
  5 & CTA 102 & 425 & 6 & 6 & 0.4060 & 10.8766 & 1.1721 & 16.0 & 3.0 & 16.5  \\ 
  6 & 3C 279 & 502 & 8 & 7 & 0.3966 & 10.9338 & 1.7201 & 16.0 & 4.5 & 20.0  \\ 
  7 & PKS 1424-418 & 473 & 7 & 9 & 0.3526 & 9.6935 & 1.2379 & 14.0 & 2.5 & 17.5 \\ 
  8 & PKS 1502+106 & 384 & 6 & 8 & 0.3326 & 17.7432 & 0.8783 & 13.5 & 8.0 & 13.0  \\ 
  9 & TON 0599 & 355 & 7 & 11 & 0.2911 & 178.3333 & 0.3183 & 10.5 & 13.0 & 7.0  \\ 
  10 & 4C+38.41 & 462 & 7 & 9 & 0.2816 & 15.1153 & 0.7943 & 11.0 & 7.5 & 13.5  \\ 
  11 & 3C 273 & 363 & 3 & 9 & 0.2808 & 273.0000 & 0.0000 & 9.5 & 15.0 & 1.0 \\ 
  12 & BL Lac & 475 & 3 & 11 & 0.2784 & 12.1855 & 0.6458 & 10.5 & 4.5 & 13.0  \\ 
  13 & PKS 0454-234 & 472 & 3 & 9 & 0.2588 & 23.2762 & 1.1288 & 8.5 & 9.5 & 15.5  \\ 
  14 & 1ES 1959+65 & 420 & 5 & 8 & 0.2318 & 330.0000 & 0.0000 & 7.0 & 16.0 & 1.0  \\ 
  15 & Mrk 421 & 509 & 6 & 10 & 0.2186 & 19.3333 & 0.1606 & 6.0 & 8.5 & 10.0 \\ 
  16 & ON+325 & 447 & 3 & 9 & 0.2164 & 357.0000 & 0.0000 & 5.5 & 17.0 & 1.0  \\ 
  17 & S5 0716+714 & 490 & 4 & 10 & 0.2029 & 47.0549 & 0.3407 & 3.5 & 11.0 & 11.0  \\ 
  18 & Mrk 501 & 461 & 2 & 8 & 0.2001 & 371.0000 & 0.0000 & 3.5 & 18.0 & 1.0\\ 
  19 & 3C 66A & 494 & 3 & 9 & 0.1886 & 404.0000 & 0.0000 & 2.0 & 19.0 & 1.0  \\ 
  20 & PKS 2155-304 & 507 & 6 & 7 & 0.1850 & 417.0000 & 0.0000 & 1.0 & 20.0 & 1.0  \\ 
   \hline
\end{tabular}
\caption{The non-linear time series analysis applied on the \gama-ray light curves of 20 blazars.
The  second column in the table shows the name of the source, third the available length of the  observations  in weeks, fourth the estimation of time delay calculated by AMI (see Section \ref{ami}), fifth is the estimation of embedding calculated by L. Cao's method (see Section \ref{Cao}). The 6th to 8th columns are  the mean RQA measures  mD, mL, and mEN, respectively; 9th to 11th are  the mean scoring measures of RQA msD, msL, and msEN, respectively. The time-lag ($\tau$) and embedding $m$ for RQA function input is taken as maximum of all the observations and is of the value of 9 and 11, respectively.
 The scoring is introduced in order to capture the RQA information across the considered thresholds. However, the msEN values corresponding to the mEN values of the same magnitude (0) are denoted by 1.0. A similar table  is presented in Appendix for light curves made evenly spaced by  linear interpolation.}
 \label{TAB1}
\end{table*}


\subsection{ Deterministic study }

Non-linear time series analysis (NLTSA) serves as a powerful apparatus which can directly probe into the dynamical states of a deterministic systems.  It also provides a framework for the inverse problem complexity, which in some cases can help regain the equations of motion of the underlying system.  In the current work, as an complementary study to \citet{Bhatta2020} and in direct contrast to stochastic modeling of the observed variability study, NLTSA is carried out on the \gama-ray light curves of 20 blazars with a deterministic approach such that the light curves are modeled as the output from the low/high order dynamical systems. As in the Fourier transformed based analyses, to characterize the deterministic properties of the astronomical observations could be a challenging task, because the methods, in principle, demand observations for an  infinite length of time \citep[see][for mathematical proof]{1981LNM...898..366T}. 

Moreover, in NLTSA, the appearance of chaotic properties of deterministic systems could be of diverse nature and depend upon a number of factors e.g., number degrees of freedom of the underlying systems, the measurement error and signal to noise  ratio \citep[see][for an overview on Chaos]{{2015Chaos..25i7610B}}.  Nevertheless, estimates of the relation between these quantities can be found, for example, in the well known invariant method of estimation of fractal dimension, the relation between the number of observations in the original state space and the correlation dimension can be constrained as $N > 42^{D_2}$ \citep{1988PhLA..133..283S}, where $D_2$ is  - correlation dimension \citep[an effective algorithm of correlation dimension can be found in][]{1983PhyD....9..189G}. In particular, Chaos analysis of scalar time series can be approached following several methods e.g. invariant methods such as fractal dimensions (box counting, correlation dimension) or numerically calculated Lyapunov exponents \citep[see][]{Kantz}.
We have demonstrated applicability of some of these methods while treating relations of the chaotic and regular motion around magnetized black holes \citep{2019EPJC...79..479P}.
From the well established NLTSA methods, we employ the Recurrence Quantification Analysis  \citep{1992PhLA..171..199Z} and the Practical method for determining the minimum embedding dimension of a scalar time series introduced by L. Cao \citep{1997PhyD..110...43C}.  These two approaches are preferable for the work as the methods are less sensitive to the gaps in the data and work well for reasonably finite number of observations. The number of observations for the $20$ blazars presented in \citep{Bhatta2020}  vary around $430$ and are reasonably evenly sampled and therefore well suited for such an analysis.

Furthermore, we also realize the importance of the choice of the parameters in obtaining most reliable results. In the context of RQA method, we adopt an approach where we consider presenting RQA as the function of thresholds instead of RQA measure for a single value of threshold.  Similar approach was  implemented in \citealt{Suk1}  for the calculation of significance of chaotic processes. The underlying assumption of this approach is that the recurrence measures are significant on different scales (thresholds). This aspect of the analysis based on Reccurence Plots (RP)  can be well observable in unthresholded reccurence plots  \citep[e.g. see RPs in][]{rproc}. Consequently, a RQA measure over  range of thresholds should be more accurate than just considering one fixed value. It exhibits more rigorous deterministic behavior in a system along with its properties on different scales.

For the estimation of the embedding in relation to the degrees of freedom of underlying system, we use the method developed by L.Cao, which is particularly well known for being not sensitive to the number of observations, this embedding dimension is later used as the input for the RQA analysis of the light curves. 
 There is emphasis given to obtain most unbiased result and for this purpose we present three tables \ref{TAB1}, \ref{TAB2} and  \ref{TAB5} of different configurations of the algorithms applied on real data, where the emphasis is given to the task of distinguishing between less and more deterministic signals present in the observations. For this purpose, the Tables of main results \ref{TAB1}, \ref{TAB2} and  \ref{TAB5} are presented in the descending order of the 4th column value, the averaged Determinism measure as described in  Section \ref{RQA} by Equation \ref{DET}.
The computation of the relevant quantities, i. e. average mutual information, L.Cao algorithm and  RQA  (see Section \ref{ami}, \ref{Cao}, \ref{RQA}, respectively)  performed using  ``NonlinearTseries"  pacakage in R \citep{Garcia}. Optimal parameters for above functions have been set up by testing the performance on artificial light curves (ALC) produced with RobPer \citep{10.18637/jss.v069.i09} library  (see Section \ref{test}). The application on real data in 3-ways has been done according the results of testing presented in Table \ref{TAB3}.   The artificial light curves with different configurations had especially different noise to signal ratios  and the values of input parameters were tuned in consideration of the ordering the signals according to their deterministic content.

\subsubsection{Time delay and the average mutual information} \label{ami}

The roots of non-linear time series analysis are bounded with the state space reconstruction.
 One can reconstruct the dynamics of a multi-dimensional non-linear system from a single time series using theoretical formulation based on mathematical theorems   for example \citep{1981LNM...898..366T}.
However the term ``reconstruct" is meant in the sense of  topological properties, which can be very useful in exploring the behavior of the underlying systems. The standard approach for state space reconstruction is the  delay coordinate embedding. The original scalar vector from the time series is simply mapped into new space, which is defined by the number of delayed dimensions.
The $m$ dimensional delayed  vector  $\vec{X}(t)$   constructed from $m$ samples of the $\vec{y}(t)$ with the delay $\tau$ is defined as:
\begin{equation}
 \vec{X}(t) =  [ \vec{y}(t),  \vec{y}(t-\tau),  \vec{y}(t-2\tau), \dots,  \vec{y}(t-(m-1)\tau)  ] 
\end{equation}

The embedding theorems require $ \tau $ to be any nonzero not necessarily a multiple of any orbit’s period.
However this is true only in case of infinite  amount of noise-free data. 
When dealing with real observations one works with finite data added with noise and the measurements errors.
In practice, the $\tau$ is significant factor when reconstructing the phase space, and if   $\tau$  is too small, the $m$ coordinates in each of these vectors are highly  correlated, and  the points from embedded dynamics are  close to the main diagonal of the reconstruction space and may not show any interesting structure. If $\tau$ is too large, the different coordinates may be not correlated and the reconstructed  attractor   may not be very similar to that of the underlying system.
    \begin{figure*}
\centering
\includegraphics[width=170mm,scale=1]{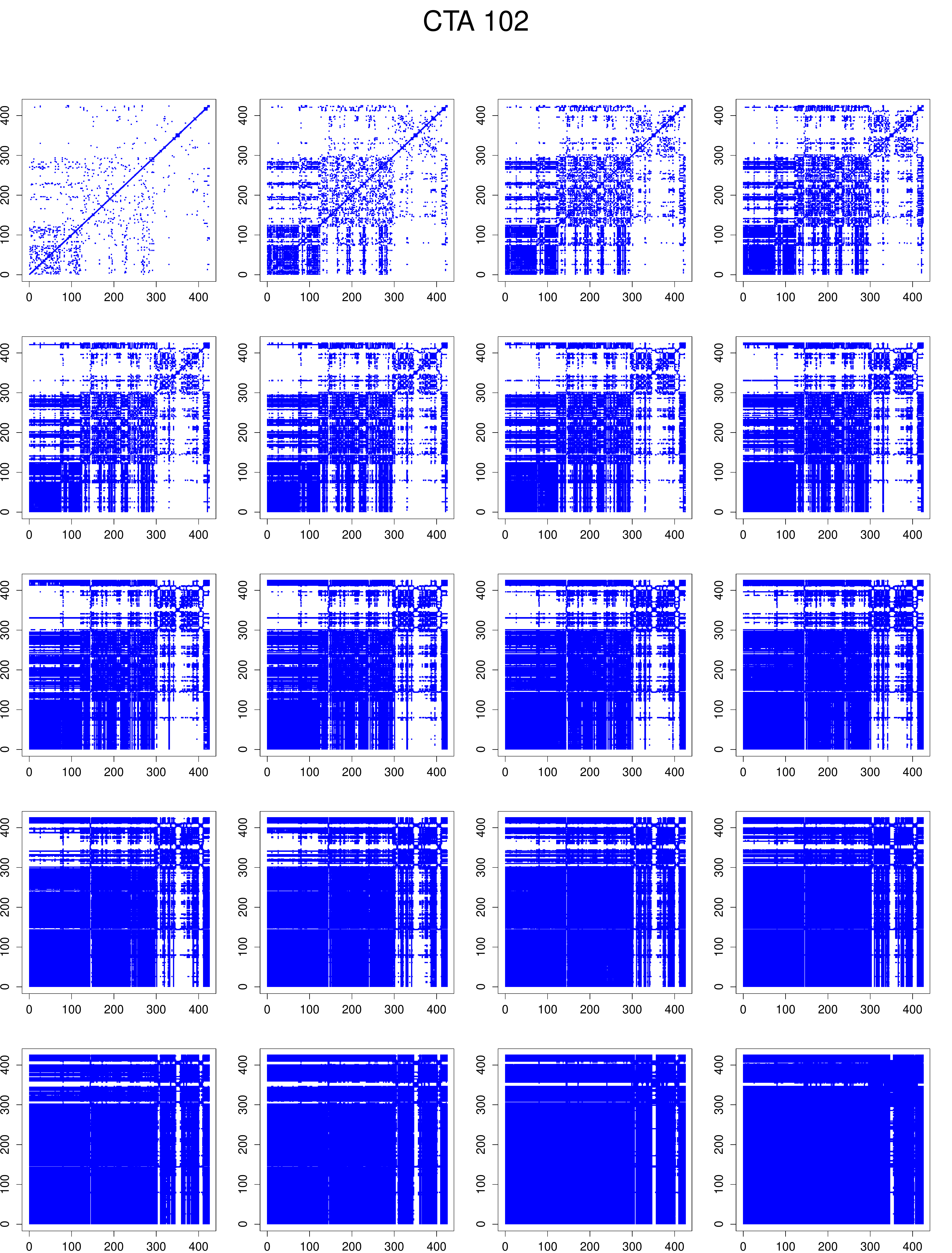} 
\caption{Recurrence plot for the source CTA 102 with various thresholds in  RR[\%] -- from 1, 5, 10, ..., 95\%.  The plots are constructed without the application of embedding.}  
\label{fig1}
\end{figure*} 

The  time delay vector defined by  $\tau$  has significant impact when comparing the results of chaoticity measures for different set of observations. This has been observed when running many simulations with artificially produced data, therefore we tested in Section \ref{test} three configurations of the $ \tau $ choice,  namely, 
a) the same value for every set of compared observations chosen as maximum or mean of the set, b)
its own value for every observation. This  $ \tau $ value, calculated by the 
Average Mutual Information (AMI; see Section \ref{ami}) was taken as maximum for RQA input in main results in Tables   \ref{TAB1} and \ref{TAB5}   for the compared sets  of light curves  of not-interpolated observations and as own value for every observation in Table \ref{TAB2} of interpolated observations.

Mutual information, a measure of the information shared between two random variables, is also used by stochastic modeling \citep{Jiao}.
In the framework of NLTSA of a observed time series  $x(t)$,  AMI denotes the amount of knowledge mined into the neighborhood of $x( t +  \tau )$.
The AMI algorithm as described in \citep{Kantz} uses the interval explored by the data  where it constructs a histogram of $\epsilon$  resolution  for the probability distribution of the data. If $p_i$ is  probability that the signal has a value in the  i-th  bin of the histogram and $p_{ij} $ is  probability
that $ x(t)$ is in the i-th bin and $x(t + \tau)$ is in j-th bin, then AMI for the given $\tau$ is written as

\begin{equation}
\rm{AMI}_{\epsilon}(\tau) =  \sum_{i,j} p_{ij}(\tau) \ln  p_{ij}(\tau) - 2 \sum_{i}  p_{i}  \ln  p_{i}.
\end{equation}

The $\tau$  is then selected by first minimum approach, that is, $\tau$ for  which AMI function reaches its first minimum.

Another method for calculating appropriate  $\tau$ is the graphical approach and the autocorrelation function (ACF).
 By graphical approach one observes the structure on the reconstructed state space. However, the graphical approach
 is definitely not suitable when dealing with a large amount of data, as it could be computationally expensive. To manually tune $\tau$ and then plot the reconstructed phase space and decide whether it is appropriate  or not is also time consuming. There are arguments against the use of ACF in the context of non-linear analysis as the method is based of linear statistics and it could omit the non-linear dynamical correlations  \citep{Kantz}. In such context, it is often stated that  the product $m * \tau$ becomes a more relevant and meaningful measure rather than the exact values of $m$ and $\tau$  \citep{2015Chaos..25i7610B}.

\subsubsection{L. Cao's Practical method for determining the minimum embedding dimension of a scalar time series} \label{Cao}

\begin{figure}
\centering
\includegraphics[width=0.9\columnwidth]{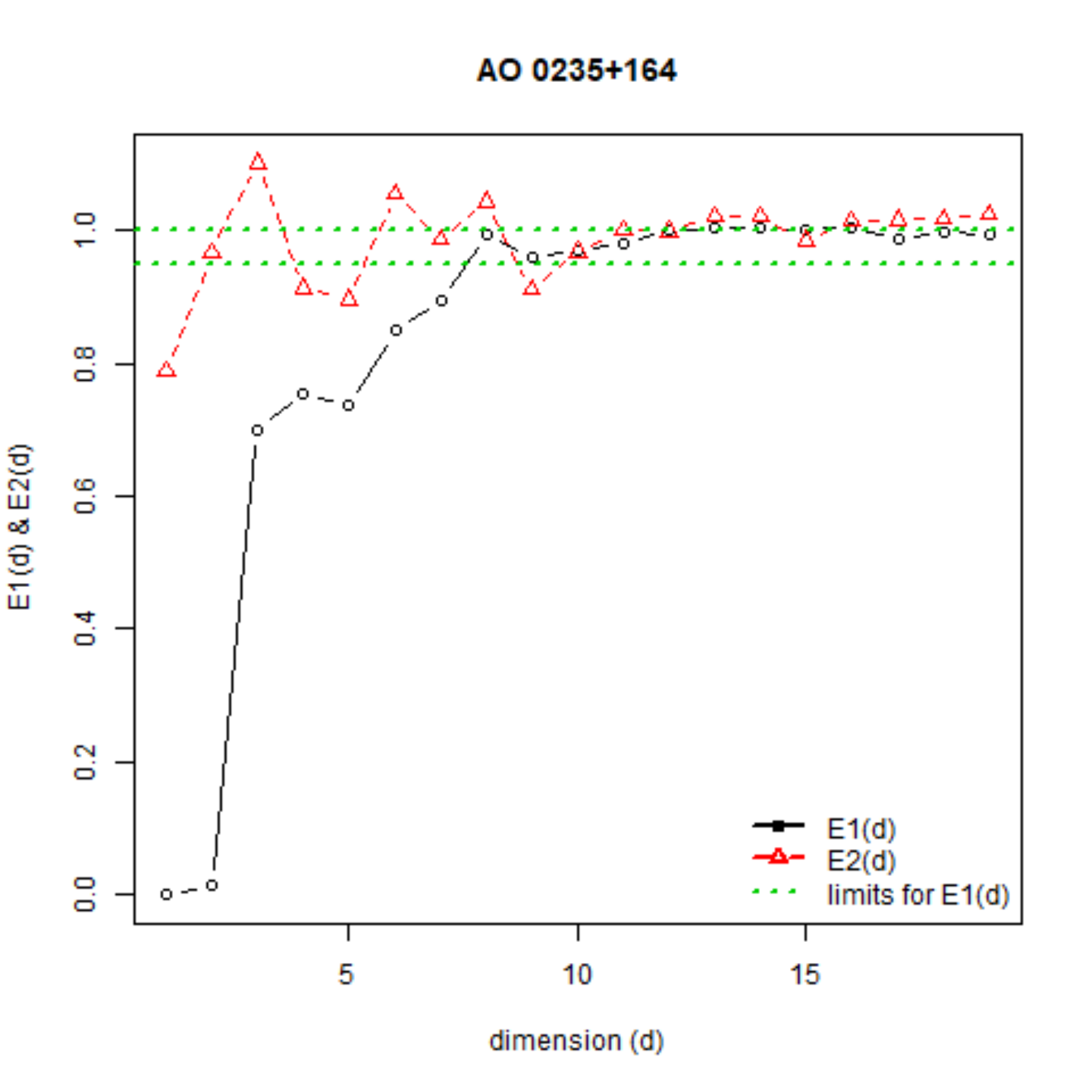}
\caption{Graphical implementation of the L. Cao's Practical method for determining the minimum embedding dimension of a scalar time series applied  the source AO 0235+164.   The red curve denotes the measure of determinism, the more the curve varies as the function of the dimension, the more determinism it contains. It is observed that the red curve for the source AO 0235+164 is varying significantly. The black curve shows the estimation of the embedding dimension.  When all of the points from some embedding  lie in the confidence interval given by green points, the given value of embedding is the estimation.}
\label{figE2}
\end{figure}

 \begin{figure*}[!t]
\centering
\includegraphics[width=1.1\textwidth,scale=1]{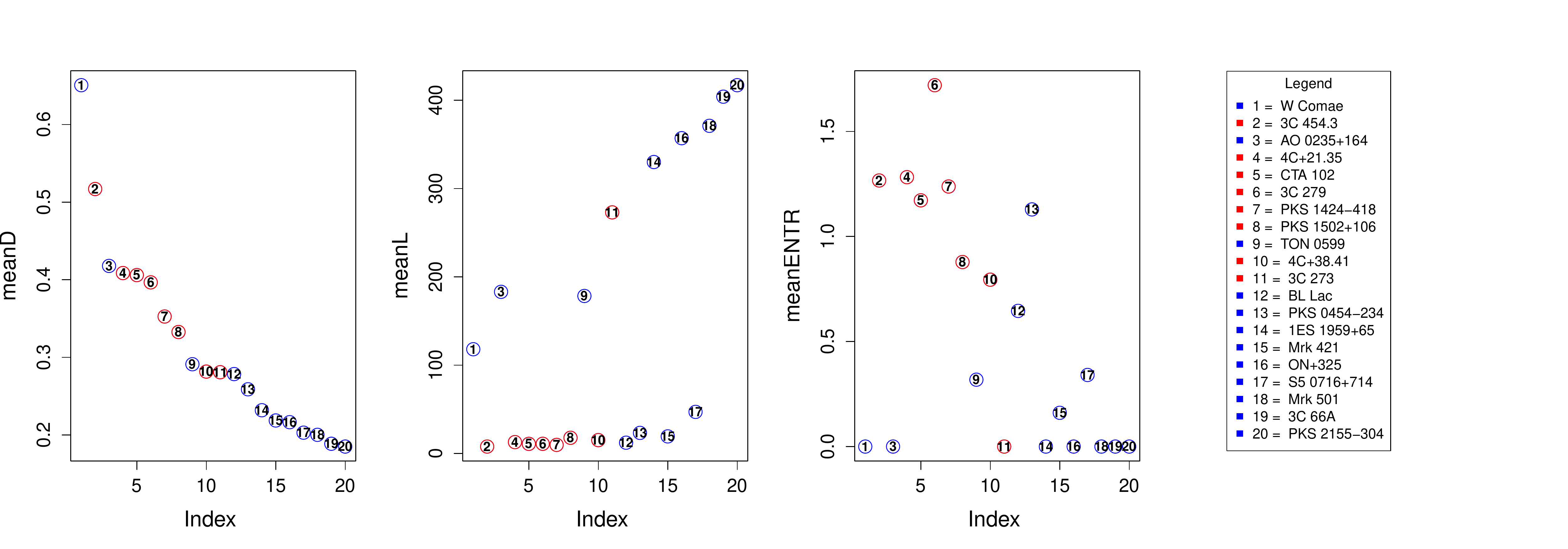}
\caption{The distribution of the averaged DET, L and ENTR as derived from the recurrance analysis and as presented in the Table \ref{TAB1} are shown in the left, middle and right panel, respectively. The sources FSRQs and BL Lacs are distinguished by the red and blue colors, respectively}
\label{fig3}
\end{figure*}

 \begin{figure*}[]
\centering
\includegraphics[width=0.8\textwidth,scale=1]{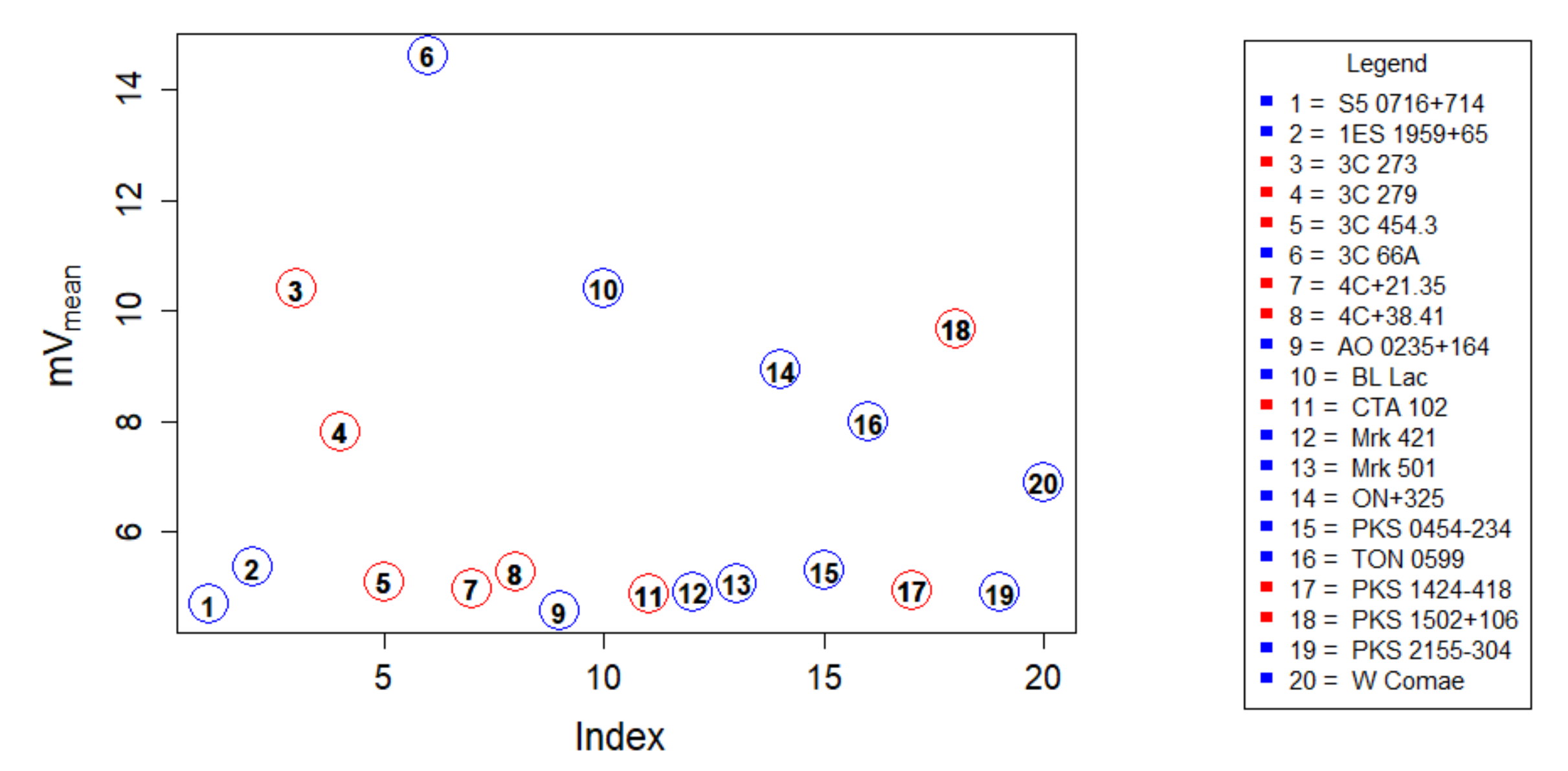}
\caption{ The characteristic dynamical timescales as represented by the average vertical line features in the recurrence plot of the sample sources along with the legend are shown here.  The RQA parameter Vmin, denoting the minimum number of points considered as vertical line,  scales by the same criteria as for the selection of lmin (see the caption of Table \ref{TAB3}). The sources FSRQs and BL Lacs are distinguished by the red and blue colors, respectively.}
\label{timescale}
\end{figure*}

This method bears several practical advantages in the estimation the minimum embedding dimension of  time series. One of them is the low number of  subjective input parameters and namely only the time delay parameter $(\tau)$. This is an big advantage performing  Chaos Analysis by nature is  strongly sensitive not only to the initial conditions but also on the number of input parameters used in modeling the data. Moreover, the implementation of the method consumes  less computational time compared to other methods e.g. invariant methods.

Let the $x_1, x_2, \dots, x_N $ be the observed data and let the reconstructed time delay vector have the form  $y_i(m) = ( x_i, x_{i+\tau}, \dots, x_{i+(m-1)\tau}) $   where $i = 1,2, \dots, N - (m-1)\tau$,
where $m$ denotes the embedding dimension and $\tau$ the time delay, then  $y_i(m)$ denotes the i-th reconstructed vector in state space with embedding dimension $m$. 
Next the variable $a(i,m)$ is defined similar as in False nearest neighbors method:

\begin{multline}
 a(i,m) = \frac{\left\|  y_i(m + 1) \: - \:  y_{n(i,m)}(m + 1)  \right\|}{\left\|  y_i(m) \:  - \:  y_{n(i,m)}(m)  \right\|} \ \ \ \\ \text{with} \ \ \  i =  1,2, \dots N - m \tau
\end{multline}

where  $n(i,m) \in (1  \leq  n(i,m)  \leq N - m \tau)$  is an integer for which $   y_{n(i,m)}(m ) $ is the nearest neighbor of  $ y_{i}(m) $ in the reconstructed $m$-dimensional state space, naturally    $  y_{n(i,m)}(m)  $  is nearest to the $ y_{i}(m) $ in some euclidean norm when  $  y_{n(i,m)}(m) = y_{i}(m) $, the nearest neighbor is the smallest  $ n(i,m)  $
for which  $   y_{n(i,m)}(m) \neq  y_{i}(m) $.
When two points are close in the $m$-dimensional reconstructed space, and  also   $(m + 1)$-dimensional reconstructed space they are called true neighbors otherwise they are  false neighbors. The feature of true neighbors comes from the embedding theorems such \citep{1981LNM...898..366T} and for a perfect embedding no false neighbors does exist. In order to omit defining some value for which  $a(i,m)$ is sufficiently small L. Cao defines the mean value $E(m) $ of all 	$a(i,m)$ values.  
\begin{equation}
E(m)  = \frac{1}{N - m \tau }  \sum_{i = 1}^{N - m \tau } a(i,m). 
\end{equation}

 To determine right $m$ the $E1(m) =  E(m)/ E(m + 1) $ is introduced and $E1(m)$ stops changing when $m$ reaches some value of $m_0$ and then $m_0 + 1 $  is the estimation of the embedding dimension \footnote{This estimation is taken as the input for the RQA.}.

\subsection{Recurrence quantification analysis \label{RQA}}
The recurrence quantification analysis (RQA) is  an apparatus  which measures the properties of the Recurrence plot (RP), a graphical tool  introduced by \cite{1987EL......4..973E} that is used for investigating the state space trajectories. 
 RQA was introduces in 1992 by \cite{1992PhLA..171..199Z} and later improved by \citet{2008EPJST.164....3M}.
In RQA, the basis for calculating RP  is provided by the matrix defined as  
\beq   
R_{i,j}= H(\epsilon - \| x_i - x_j\|  \mid ) \quad  i, j = 1, ...,N,   \label{Eq:1}
\eeq
where $N$ is the number of measured points $x_i$, $ \| \cdot \|$ is a norm and $\epsilon$ is a threshold distance which is crucial value which has a strong effect on the result.  H(·), the Heaviside
function, is defined as 

\begin{equation}
H(\epsilon) =
    \begin{cases}
            0, &         \epsilon < 0\\
            1, &     \epsilon \geq  0.
    \end{cases}
    \label{Eqn4}
\end{equation}
 It can be seen that Equation \ref{Eq:1} gives rise to a symmetrical square matrix that consists of binary values, i. e., zeroes and ones. The RP is obtained as a plot of this square matrix. As threshold value parameter largely determines density of the RP plot, there appears to be some ambiguity over a consistent choice of  the $\epsilon$   \citep{2008EPJST.164...45S}. Therefore instead of
  looking for one single set of correct values, a more rigorous result could be obtained by averaging over more thresholds.

  {In this work, we follow the approach in the implementation of the RQA method by  averaging over the span of thresholds, where in our case the thresholds are calculated for  a wide range of percentages of points in RP (ones in the binary matrix) and thereby consider the significance of RQA measures on diverge scales.  This percentage is actually part of the RQA analysis defined as  the recurrence rate (RR) 
 \beq
RR = \frac{1}{N^2} \sum_{i,j=1}^N R_{i,j}    \label{RR},
 \eeq
 which provides a measure for the density of the recurrence points in the RP.   As an example case, the RPs for the blazar CTA 102 are shown in the panels of Figure \ref{fig1}.  Dictated by Eqn. \ref{Eqn4}, the panels shows that as the RR is increased from 1-95\%  with an step of 5\%, i.e. [1, 5, 10, \dots, 95]\%,  the area in the plot is gradually populated by larger number of blue symbols.

Determinism - Determinism is computed considering the RR of the points which align along the diagonal lines of the RP. The quantity tells how deterministic or well behaved a system is. The mean determinism, over the range of the RR[\%] considered here, of the sample blazar \gama-ray light curves, denoted as ``mD", are presented in the 4th column of the Table \ref{TAB1} and \ref{TAB2}, and note that all the other columns values in the tables are sorted according to descending order of the mean determinism values.
\beq
\rm{DET}  = \frac{\sum_{l=l_{min}}^N l P(l)}{ \sum_{i,j=1}^N R_{i,j}}, \label{DET}
\eeq
where  $P(l)$ denotes the frequency distribution of the lengths $l$ of the diagonal lines.
\\
L - Line length, represents average diagonal line length in the RP and directly related to the predictability time of the system. In the context of the light curves, the quantity mirrors the average time during  which any two  flux points are close to each other. This time can be interpreted as mean prediction time 
\beq
L = \frac{\sum_{l=l_{min}}^N l P(l)}{\sum_{l=l_{min}}^N  P(l)}.
\eeq
The mean line length, over the range of the RR[\%] considered here, of the sample blazar \gama-ray light curves, denoted as ``mL", are presented in the 5th column of the Table \ref{TAB1},\ref{TAB2} and \ref{TAB5}.
\\

 ENTR - Entropy, computed as the probability distribution of the diagonal line of lengths p(l) of the RP, provides a measure of the complexity of the data. In other words, the quantity conveys how much information the observation does contain -- or richness of the data.
\beq
   \rm{ENTR} =   - \sum_{l=l_{min}}^N  p(l) \ln p(l),
\eeq
where $p(l)$ is the probability that a diagonal line in the RP is exactly of the length $l$ -- it can be estimated from the frequency distribution $P(l)$ with\\
 $p(l) = \frac{P(l)}{\sum_{l=l_{min}}^N  P(l)}$.   The mean entropy, over the range of the RR[\%] considered here, of the sample blazar \gama-ray light curves, denoted as ``mEN", are presented in the 6th column of the Table \ref{TAB1}, \ref{TAB2} and \ref{TAB5}.
 \\
In the computational process,  for every RR[\%]  $\in$ [1, 2] in Table \ref{TAB1}, RR[\%]  $\in$ [1, 2, 3] in Table \ref{TAB2} and  RR[\%] $\in$ [1, 5, 10, \dots, 95] in Table \ref{TAB5},  RQA measures (D,  L, and EN) were calculated for the observations and averaged into  mD, mL, mEN, as well as they were listed in descending order of  D, L, and EN, sequentially, and then integers from 1-20 (number of analyzed observations)
were assigned to them as sD, sL, and sEN,  respectively. One can visualize the scoring  sD, sL, and sEN measures (for some RR and all the observations) as an column with values of $(20, 19, \dots, 1)$ aligned in descending  order   D, L, and EN measures (20 for highest value of D, L, and EN, 19 for second highest etc.).

The additional statistics of averaged scoring measures msD, msL, and msEN of sD, sL, and sEN provide additional information about times the observation scored at some place for considered percentage of RR. The ordering of averaged scoring can be different from the averaged RQA measures mD, mL, and mEN as it can seen that when comparing mD and msD columns. This means that a source that is most deterministic by mD measure might have slightly less deterministic scoring.  
For the sample the averaged RQA quantities  DET, L and ENTR  for RR[\%] $\in$ [1, 2] are presented in the Table \ref{TAB1} (denoted by  mD, mL, mEN) and are also depicted in Figure \ref{fig3}. From the Table and the Figure, it can be seen  that the averaged RQA measure DET is relatively large for some the sources in the sample and especially for FSRQs. This implies  dominant role of the deterministic processes that lead to the observed \gama-ray variability in the sources. 
Moreover, Figure \ref{fig3} shows an interesting features that, the RQA measures on average as  FSRQs have larger values of  DET, L and ENTR compared to BL Lacs. This pattern is also to observe in Table \ref{TAB5} with RR[\%]  $\in$ [1, 5, 10, \dots, 95] so as in  \ref{TAB2} of linearly interpolated observations, where  RR[\%] $\in$ [1, 2, 3] with different orders. We also note that the analysis results zero mean Entropy for 8 of the sources, mostly BL Lacs. This possibly could have been caused by the relative `low information content'  in the light curves of these sources.

As non-linear phenomena and chaotic systems are sensitive to initial conditions, the non-linear methods and algorithms are also sensitive to the inputs. Therefore, it is not surprising that the results presented in Tables \ref{TAB1},  \ref{TAB2} and \ref{TAB5} differ in the orders and the magnitudes of the measures.
However the difference in Tables \ref{TAB1},  \ref{TAB2} and \ref{TAB5} is more significant in magnitude of the RQA measures than in the order of sources according their deterministic content, therefore the rather  than to make comparison of the magnitudes of the RQA measures among the tables, the comparison of the RQA measures of the sources within single table  makes more sense.
Figure \ref{fig3} shows a the pattern of FSRQs with higher deterministic content, which is also observable from the Tables.
However, it is important to note that, the position of  blazars Wcomae and AO 0235+164 can be 
questionable when taking into account of their least number of observations - around the half of the mean length (see second column in Table \ref{TAB1}), while the length is important factor when handling non-linear phenomena from both theoretical and practical points.

In such context, the source blazar 3C454.3  can be seen as one with highest averaged deterministic value according to Table \ref{TAB1}; this blazar is second in the Table \ref{TAB2} behind the 4C+21.35  and in the Table \ref{TAB5} the FSRQ PKS 1502+106 has the most deterministic content.

In  Figure \ref{fig3} we observe that in  mEN measure the FSRQs also correlate with the mD measure and show more information content than BL Lacs. Consequently, the sources are less predictable according to mL measure, following from the fact that, the complex  non-linear and chaotic systems are likely to be less predictable.

In Table \ref{TAB5},  where the RQA measures are averaged to very high percentage of RR,  the distinction between FSRQs and BL Lacs is also observed.  In this case, where  RR $\in$ [1, 5, 10, \dots, 95]\%, the order of sources  would almost not change if the averaging would be set just until 50 \%, while the higher the limit of averaging is set up the bigger is the gain of mD and mEN measures.
Overall, while the mD and mEN measure in this case gained some values, the mL column is lower in comparison with Table \ref{TAB1} and \ref{TAB2}. The most significant difference in this case is the bottom of the Table showing higher values of mEN which could be explained by the ``tangential motion" phenomena occurring when the threshold values (corresponding to  high RRs) are too high, as described in \citep{2007PhR...438..237M}.
}

\subsection{Estimation of timescales: diagonal and vertical features \label{RQA}}

The RQA measures can be exploited to delve into the dynamical timescales inherent in a time series. In particular, the timescales can be computed using the line features that are parallel to the LOI and the vertically re-occurring features. The timescales based on the structures that are parallel to the LOI in the RP plot-  every parallel line represents delay times.  This tells how frequently the system re-visits the same state. The timescale estimated this way also provides a measure of auto-correlation and  is capable of revealing the quasi-periodic oscillations.  Similarly, the timescale estimated using the vertical features provide an estimation for the time   the system spends in a dynamical state; in the context of the \gama-ray light curves of blazar, the states represent particular state which emits a given amount of \gama-ray flux in the light curve \citep[see][]{2020arXiv200102800P}.

 To compute such timescales, the  observations were interpolated so that the observations in the light curves are evenly spaced. In the most of the cases, the light curves are more that 90\% evenly spaced, so the interpolation should not change the results drastically. However, in case of unevenly spaced light curves, the interpretation of the timescales represented by the delay time is not straight forward.  The RPs were constructed for the sample sources using RR[\%] from   1-3, and the recurring features, e. g. lengths of the vertical lines and time delays between diagonal lines, were computed. However, comparing different configurations of the setup of RQA, the diagonal timescales are too dependent on the setup of RQA parameters and therefore they are not presented in this work. The average of the timescales corresponding to the range of the RR[\%] was taken as the dominant timescale in the light curves. The resulting vertical timescales in weeks (y-axis in the plot) for the sample sources are shown in Figure \ref{timescale}. It seen that  trapping timescales, which reflect on stability/instability of a system, is in the order of 5-15 weeks.

 \section{Discussion} \label{sec:4}

The non-linear time series analysis performed on the Fermi/LAT light curves a sample of 20 blazars has revealed interesting results. In this section, we present discussion on the possible interpretation of the results in the context of currently accepted blazar models.

As known, the measurements by sensors do not have smooth time bases. The  embedding theorems require evenly spaced observations. This is definitely less luxurious demand as the infinite amount of observations and one can achieve this by the good known technique of interpolation. So in order to obtain most unbiased results we apply the non-linear analysis also on the interpolated data.
 We  compare the results obtained with the ``raw" data in Table \ref{TAB1},   so as the interpolated ones in Table \ref{TAB2}   and try to judge the effect of added interpolated dynamics.

When providing RQA the choice of the threshold is crucial, the recommendations for its choice have been given by many authors and most of them are derived from the variance of the observed data \citep[see][]{2008EPJST.164...45S}.
As  mentioned earlier,  for a more robust result, instead of one single value of the threshold, the RQA measures averaged (see Section \ref{RQA}) over a range of thresholds are preferred. This approach reflects the behavior of the underlying system in diverge scales, providing  more objectivity to the results. One of the important parameters of this approach is the interval of percentages one considers for this averaging. In this work, many setups of RQA are massively  tested  on artificial data in the sense of choice of minimal diagonal length - lmin, time delay - $\tau$, embedding dimension - $m$ and the value of reccurance rate - RR  where to average (see Section \ref{test}) by the search to  find best parameters for the real data of 20 blazars.

In order to present most unbiased results possible, three different  approaches to the RQA are performed on the real data, which were based on the results from the extensive testing processes using similar artificial data as presented in Table \ref{TAB3}. For every RR[\%] in the first column, 90 different setups of RQA  were performed (10 different ways of the lmin, 3 ways of the $\tau$ and 3 ways for $m$).  With an aim to configure the most suitable parameters, i. e., lmin, $\tau$ and $m$, for every RR[\%], the optimal setup was selected based on the ability  to sort the signals in the ALCs according their deterministic strengths. (For details on the testing process refer to Appendix \ref{test})
  Every setup was tested on  5 sets of 10 artificial light curves  generated using 5 different configurations of the generator:  in first three ways light curves were generated using the same configuration for all 10 different SNR ratios and in the other two ways  all of 10 LCs with different SNR ratios  had randomized  parameters. On top of these, the ALCs were provided with high red noise content to account for  the observed power-law shape of the power spectral  density of the \gama-ray observations.
Next, in order to mimic our condition the most, the gaps in the data were made.  From fixed length of the generated data set up to 513, the random amount of data up to 50\% was  replaced by NaN values. The artificial data were then also linearly  interpolated making in common two huge sets of artificial data.

From Table \ref{TAB3}, two approaches are chosen and presented on ``raw" data:  a) the approach where $ \tau = 9 $ and $m  = 11$ are set up as maximum values from the set of 20 blazars, while it is averaged  up to RR = 2\% only by the step of 1\%  (see Table \ref{TAB2}) b) the setup of maximal $m $ and the corresponding $\tau $  for every source, where in order to include the effects of a wide range of threshold,   large range of $\epsilon$  is included  in terms of RR between 1--95\% with the step of   $5\%$ (see Table \ref{TAB5})  and then average the measures over all $\epsilon$ to obtain the final results. It is noted that this setup was found to produce quite stable results when considering different inputs of $\tau$ and $m$.
         Although the weekly binned decade-long \gama-ray observations are fairly well sampled in terms of number of observations, they are not strictly evenly sampled. Therefore, to assess the possible effects of uneven sampling of the light curves on the analysis,  the light curves were made evenly spaced by linear interpolation and  the analysis is performed in the similarly way. To illustrate the data interpolation, the real and interpolated observations for the two blazars, namely 1ES 1959+65 and TON 0599 are shown Figure \ref{ILC}. The corresponding quantities resulted from the analysis are presented in Table \ref{TAB2} in Appendix. 
 We note that there are no significant changes in terms of the measure RQA quantities in comparison with Table \ref{TAB1}, but now as more points due interpolation is introduced so as the RR is averaged slightly higher by 1 \% the mEN gained some values. 
 The setup chosen for interpolated data has also maximal value of $m$ and own value of $\tau$ for RQA input and it was averaged until 3\% by the step of 1\%.
 The precision was set up to one hundredth for RR $ \leq 5 \% $  and  one tenth for RR $ > 5 \% $.

The processes in AGN system could consist of both deterministic and stochastic nature. Nevertheless, both log-normal flux distribution and logarithmically increasing variability as reported in \citet{Bhatta2020} present evidence that the accretion disk related modulations still dominate over the large spatial and temporal extension providing an overall nature of the variability as deterministic.  
\citet{Kiehlmann2016} in their study of EVPA rotations in the blazar 3C 279 came to the conclusion that while the low-brightness states in the blazar could be a result of a stochastic process  and most of the high amplitude variability should be the result of the underlying deterministic processes.  The results point to the strong coupling between the accretion and jet in the sense that the disk and the jet interact in a co-ordinated manner such that information about the disk process remain intact while they propagate into the jets.  In blazar systems this also could have an implication that  although the observed flux is largely dominated by the non-thermal emission from the jets, variability probes employing suitable time series analysis could still reveal  the origin of variability phenomenon to the accretion disk.

 The results that the dominant physical processes in FSRQs are more of deterministic nature  can be  interpreted  in the widely accepted scenario that jets are powered through the extraction of the rotational energy of supermassive Kerr black hole surrounded  by magnetically arrested accretion disk.  It is also possible that FSRQs are disk dominated  showing features of more powerful accretion disk. Their jets possibly  are less magnetized and consequently provide less favorable conditions to stochastic process, e. g., rampant shock and/or magnetic reconection events. Whereas BL Lacs jet have been found to be abundant with streaming particles that can contribute to the enhanced stochasticity \citep{Zhang2014ApJ}.

  Alternatively,  the deterministic features of high energy emission can be linked to the shock compression events in  ordered magnetic field in a axisymmetric linear jets such that the turbulent activities are less prevalent \citep[e. g. see][]{Aller2020,Zhang2015}. The geometry of the magnetic field of the blazar jets have been routinely explored using multi-frequency polarimeters (e. g. optical band; \citealt{Blinov2016} and radio band; \citealt{Anderson2019}). In particular,  highly polarized  jets are indicators of large ambient jet magnetic fields. In addition, the sudden electric vector position angle (EVPA) rotations \citep[e. g.][]{Marscher2008} have been routinely observed. 
 Such EVPA rotations could be indicative of the deterministic process \citep[see the discussion in][]{Kiehlmann2016} which can be related to the strongest \gama -ray flares  frequently observed  in blazars (e. g. \citealt{Abdo2010nat,Blinov2015}; also see the \gama-ray light curves of the blazars presented in \citealt{Bhatta2020}). However, the role stochastic processes \citep[e. g.][]{Marscher14,bhatta13,Lehto1989,Jones1985} can not be completely ruled out. 
 
 Similarly, the presence of circum-nuclear material, as suggested by relatively stronger emission lines in FSRQs, that are believed to provide the low energy photons for the inverse-Compton process (External Compton model), giving rise to dominant \gama-ray emission, make such systems more complex; whereas self-Compton origin of the high energy emission (Synchrotron Self-Compton model), involving less number of interacting components, e. g. electron density distribution, magnetic field, and  Doppler factor, could imply relatively simpler scenario.

As seen in Figure \ref{timescale}, the timescales are derived from the vertical  distribution  of the points in the RP are  in range $\sim$ 5 - 15 weeks.   The timescales in the former case are indication of the average recurrent timescale of the dynamical processes signifying how frequently the system revisits a particular state. Moreover, it is interesting to note that the average predictability timescales is comparable to the recurrent timescale. It should be noted that the flux distribution being log-normal, the light curves are dominated by lower fluxes, and therefore it is natural to expect the  trapping timescales to be in the order of a few weeks.  This means the resulted average dynamical timescales represent low flux level fast variability, giving lower weight to large flares lasting several months. Nevertheless the resulting timescales could be relativistically dilated through the relation $R \gtrsim \delta t/(1+z)$, reflecting the cosmic expansion. For an average red shift of $z=0.6$, and for a moderate value of Doppler factor $\delta=10$, the timescales can be translated into size of the regions following causality argument. For a typical black hole mass of $10^9\ M_{\odot}$ with gravitational radius R$_g=GM/c^2$, the size corresponding to 15 weeks corresponds to 0.5 pc, which is comparable to the size of the inner accretion disk; and 10 weeks represents a few thousands of gravitational radii, within which most of the gravitational potential energy is converted into the radiation energy. In such interpretation, the inner accretion disk might be treated as the main component of a dynamical state of an AGN, and the modulations driven by various  instabilities e.g.  radiation pressure, viscous instabilities \citep{Janiuk2011,Janiuk2002} occurring within this region leading to the change in the states.

 \section{Conclusions} \label{sec:5}
 We probed a sample of 20 blazars   by performing non-linear time series analysis of their  decade-long \gama-ray light curves from the Fermi/LAT telescope. The results of the analysis suggest that the dynamical processes responsible for the \gama-ray variability of the blazars are mostly a mixture of deterministic and stochastic in nature, although in some of the sources e. g. blazar 3C 454.3, 4C+21.35 and CTA 102, displayed high deterministic content. The result could be  significantly useful in formulating the model that explain the interplay between the disk and jet processes ubiquitous in black holes systems. In addition, the analysis reveals characteristic timescales in a several weeks, which could be interpreted as so called  trapping timescales  ($\sim$ 5-15 weeks). The timescales in combination with the results from the multi-frequency studies could provide further insights about the nature of origin of the \gama-ray in radio-loud jets. 

\newpage

\acknowledgments

GB acknowledges the financial support by the Narodowe Centrum Nauki (NCN) grant UMO-2017/26/D/ST9/01178.
RP thanks to the SGF program CZ$.02.2.69/0.0/0.0/19\_073/0016951$  of the Silesian University in Opava. ZS thanks for the support of the Institute of Physics and Research Centre of Theoretical Physics and Astrophysics, at the Silesian University in Opava.
 We are very thankful to the anonymous referee for his/her thorough and careful reading of the paper and very useful comments and suggestions which helped improve the presentation of the paper significantly. This research has made use of the NASA/IPAC Extragalactic Database (NED) which is operated by the California Institute of Technology, under contract with the National Aeronautics and Space Administration.

\bibliographystyle{aasjournal}

\appendix
\restartappendixnumbering 

\section{RQA of the interpolated blazar light curves}

\begin{table*}[ht]
\centering
\begin{tabular}{llrrrrrrrrr}
  \hline
 & source & Len & $\tau$ & m & mD & mL & mEN & msD & msL & msEN  \\ 
  \hline
1 & 4C+21.35 & 512 & 4 & 8 & 0.5993 & 15.6162 & 2.5014 & 19.00 & 8.67 & 17.33  \\ 
  2 & 3C 454.3 & 513 & 9 & 9 & 0.5619 & 7.6438 & 1.4513 & 18.00 & 1.00 & 11.00  \\ 
  3 & CTA 102 & 513 & 10 & 9 & 0.5109 & 10.1990 & 1.7812 & 18.67 & 3.67 & 13.67  \\ 
  4 & 3C 279 & 514 & 8 & 8 & 0.4778 & 10.9323 & 2.0061 & 17.67 & 4.67 & 16.00 \\ 
  5 & PKS 0454-234 & 515 & 3 & 9 & 0.3730 & 21.7236 & 2.5954 & 14.67 & 11.00 & 19.33  \\ 
  6 & AO 0235+164 & 511 & 17 & 6 & 0.3528 & 9.0909 & 1.0025 & 14.00 & 3.33 & 8.00  \\ 
  7 & PKS 1502+106 & 515 & 8 & 9 & 0.3413 & 10.9701 & 1.6412 & 14.33 & 5.67 & 13.00 \\ 
  8 & W Comae & 509 & 4 & 10 & 0.3254 & 15.3842 & 2.1845 & 13.33 & 9.67 & 17.00 \\ 
  9 & S5 0716+714 & 515 & 4 & 8 & 0.3217 & 18.0873 & 2.0249 & 12.67 & 10.33 & 14.33 \\ 
  10 & PKS 1424-418 & 506 & 13 & 7 & 0.3037 & 12.5349 & 1.5297 & 12.33 & 6.67 & 10.33  \\ 
  11 & 3C 273 & 513 & 5 & 10 & 0.2075 & 17.4184 & 1.5992 & 9.33 & 9.67 & 12.00 \\ 
  12 & 4C+38.41 & 513 & 7 & 10 & 0.2044 & 13.8741 & 0.7189 & 9.33 & 8.00 & 4.33 \\ 
  13 & TON 0599 & 513 & 9 & 8 & 0.1729 & 32.5414 & 1.0351 & 6.00 & 14.67 & 6.67  \\ 
  14 & PKS 2155-304 & 515 & 6 & 10 & 0.1699 & 18.1636 & 0.5620 & 6.33 & 9.67 & 5.00  \\ 
  15 & 3C 66A & 513 & 3 & 8 & 0.1698 & 47.9726 & 1.8597 & 7.00 & 16.00 & 15.00 \\ 
  16 & Mrk 501 & 513 & 3 & 9 & 0.1578 & 77.5414 & 1.6060 & 4.67 & 17.00 & 11.67  \\ 
  17 & Mrk 421 & 515 & 6 & 8 & 0.1443 & 59.4289 & 0.7013 & 4.00 & 16.67 & 5.00  \\ 
  18 & BL Lac & 514 & 5 & 10 & 0.1440 & 184.2397 & 0.1867 & 4.33 & 16.00 & 1.67  \\ 
  19 & 1ES 1959+65 & 514 & 4 & 11 & 0.1361 & 102.9524 & 0.9813 & 2.67 & 18.33 & 6.33 \\ 
  20 & ON+325 & 513 & 4 & 9 & 0.1310 & 368.5556 & 0.2122 & 1.67 & 19.33 & 2.33 \\ 
   \hline
\end{tabular}
\caption{The non-linear time series analysis applied on 20 blazars with linearly interpolated values, where the columns have the same meaning as in  Table \ref{TAB1}. It can be observed that in comparison with Table \ref{TAB1}, the lengths in second named column are increased, as the missing values (or ``NaN" ) were replaced by interpolation. In this analysis, every source has its own $\tau$ for the input of RQA and $m$ is taken again as maximum of all the $m$s calculated, while this choice is made according the results in Table \ref{TAB3}. For this setup, the distinction between FSRQs and BL Lacs, in terms of deterministic content as represented by mD, is also significant. }
 \label{TAB2}
\end{table*}

\section{Testing process  }  \label{test}

In order to configure the optimal set of the parameters to be employed in the analysis on the real observations, several tests  on artificial data were performed prior to the application of averaged RQA analysis on data set of decade-long \gama-ray light curves of 20 blazars. The tests made use of several time series algorithms that deal with nonlinear phenomena. The choice of the methods used in the analysis are based on the two main criteria: First, the methods should involve minimal amount of inputs, this is crucial because when dealing with nonlinear phenomena such as chaos, the results are very sensitive to the inputs  of the algorithms. In addition, the usage of low number of inputs has several benefits e. g. the analysis can be performed for a number of combinations of the inputs within computational resources, and the interpretation of the results is easier in terms of physical theories -- in contrast to the machine learning algorithms that can provide better fits, however, the interpretability, in some sense, is often  in terms of a black box.
These tests are carried out  in order to select suitable parameters from bounded  parameter space for estimating  embedding dimension, time delay, and optimal recurrent rate, that subsequently can be fed to  AMI (Section \ref{ami}), L. Cao algorithm (Section \ref{Cao}), and RQA (Section \ref{RQA})  analysis.  The  RQA measure can mathematically be described in terms  of a compound function such as :
 \beq RQA(  \epsilon ) = RQA( lmin ( RR( \epsilon ) ) ),
  \eeq
where $lmin$ is the parameter for calculation the diagonal features in RP, which defines how minimally many diagonally connected points are considered as a line, RR is the reccurance rate defined by equation \ref{RR} and $\epsilon $ is the threshold value see Eqn. \ref{Eq:1}, which simply says what is the distance between points in order to denote it by 1 by Heaviside function and latter denote by color (not white in RP). 

The approach of the testing i. e., search for the optimal parameter setup,  is motivated to distinguish deterministic signals  from stochastic noise.  In the context of RQA,  the strength of the deterministic part of the signal is represented by the DET measure (Section \ref{DET}).  
The R package RobPer allows to generate artificial light curves with different configurations, while one of the parameters is the strength of the signal,  where the power law noise is generated according to the prescription discussed in \citet{TK}.

\begin{figure*}
\centering
\includegraphics[width=170mm,scale=1.15]{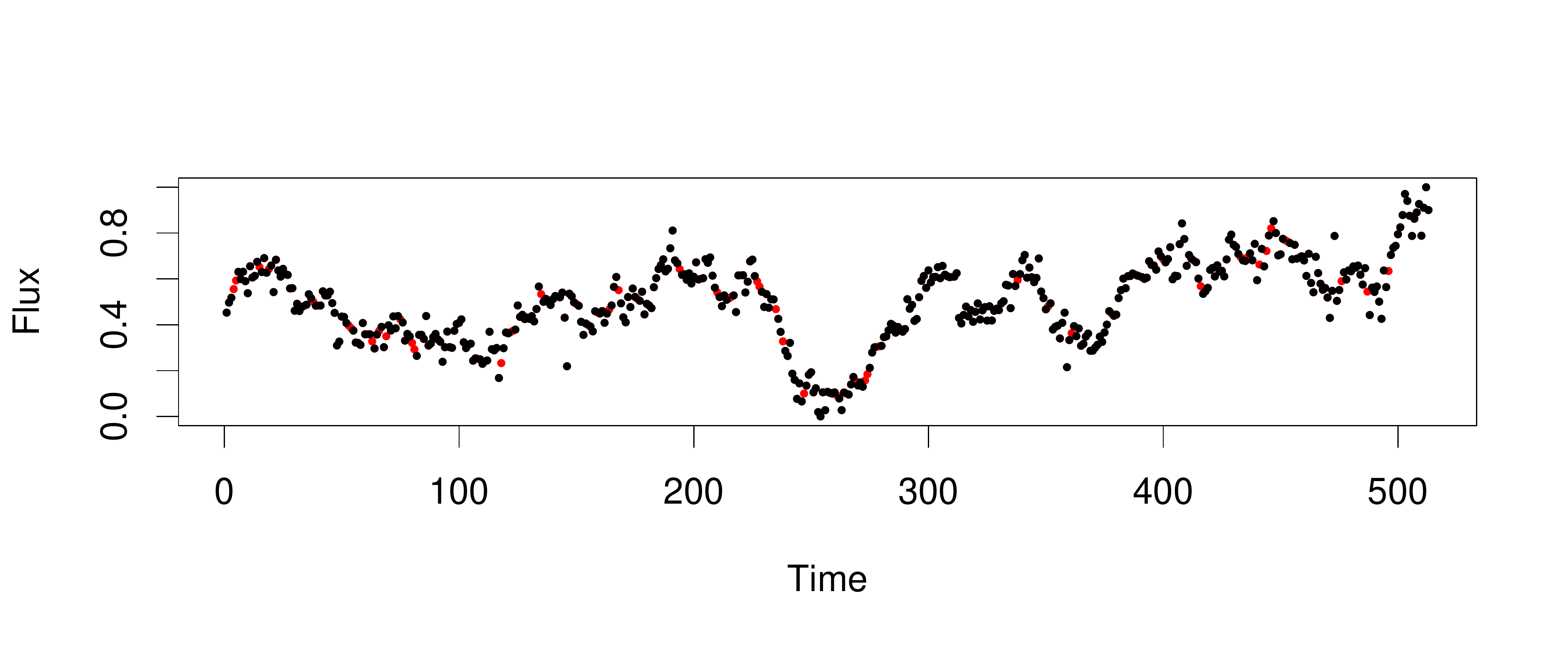}
\caption{An example of artificial light curve generated using the \textit{tsgen} function from the R package RobPer \citep{10.18637/jss.v069.i09}, which was used  to configure  the most suitable RQA parameters  and later applied on real data. The red points denote the linearly interpolated values which were deleted in order to mimic the sampling of the blazar \gama-ray light curves. }
\label{ALC}
\end{figure*}

For the training purpose, the parameters for 10 artificial light curves (ALC) were configured for several values of signal to noise ratio (see SNR from tsgen function in  \citep{10.18637/jss.v069.i09}) 
 taking values from the vector [0.005, 0.01, 0.025, 0.5, 0.75, 1 ,1.5, 2, 3, 5]. In order to produce most unbiased result on the data of 20 blazars the setup is applied on both interpolated and not interpolated data.   The  analysis were performed on artificial data which mimicked the sampling of the real observations (see Figure \ref{ALC}).  From the generated ALCs of the length 513 ($\sim$ 10 years) were randomly deleted values from the observations up to 50\%, where this number has been given according the shortest data length belonging W Comae. 

To encapsulate the RQA behavior of the varied nature of the light curves, for each of 5 different configurations  10 ALCs with different SNR ratios and noise content were generated (see Section \ref{sec:4}). In addition, the ALCs were provided  10 additional different parameters including length of the observations (see tsgen   \citet{10.18637/jss.v069.i09} for details). During the process, for the 3 out of 5 configurations  these parameters were fixed for every SNR ratio, whereas for the other two configurations the most of  the parameters (for given boundaries) were randomized. In addition, the ALCs had high red noise ratio,  up to 95 \%, in the white noise/red noise mixture. In case of the first  3 configurations  for generating ALCs, as for the selection of the main inputs to the RQA namely, time delay - $\tau$, embedding dimension - $m$ and the choice of the minimum line length -  lmin,   approaches based on multiple criteria were adopted.
The values of  $\tau$ and $m$ have been considered for testing for the set of ALCs as a)  specific to each ALC, b) the mean of the whole considered set c) the maximal value from the considered set.  So there are 9 ways to configure the tau/$m$ setting for the RQA. Lmin playing one of the key roles in RQA,  also in the sense of the magnitude of the RQA measures.  In many computational libraries, e. g.  ``NonlinearTseries", ``RHRV", and ``crqa",   this value is pre-defined  as 2 or 3. In our parameter space, lmin is  configured in 10 ways, while 4 of them are of the fixed value of  2,3,4,5 and the rest is changing value of lmin according to the RR value which is used. Changing lmin value with higher RR seems like a natural approach which can avoid ``tangential motion" which appears when the RR –   $\epsilon$, value is higher \cite{2007PhR...438..237M}. In \citep{1986PhRvA..34.2427T}, the authors recommend that for calculation of correlation integral the choice of lmin be similar to the choice of the Theiler window. However, the configuration of lmin according to RR have not been explored much and  could be an active field of  enormous possibilities. We considered 6 approaches of choosing lmin according to RR  in Table \ref{TAB3}.

All the computation has been performed on Intel TM core i7 processor of 7-th generation. The computation is mostly demanding in sense
of many times repeating (looping) the computation of RQA in order to find the desired threshold ($\epsilon$) for given RR. The
computation of RQA could have not been improved by using available R libraries which use GPUs, while the efficiency
of GPU comes with long data sets. In frames of the computational resources the precision of finding threshold for RR
has been within tolerance of 5 hundredths for RR lower than 5 \% and 5 tenths is equal or greater 5 \%.

To summarize, in order to parameter setup and search the optimal parameters, an extensive test was performed using a large number of ALCs, i. e.,  on each   group  of  10 ALCs with 5  different settings (divided into two groups - with gaps and interpolated ones) RQA algorithm has been run in loop while desired threshold  for given RR  with defined precision  have been found,  while this process have been repeated 9 times for a combination of 3  $\tau$s and 3 $m$s. After the desired threshold for RR has been found, 10 different approaches of choosing lmin has been computed. Eventually, on each of 100 ALCs, 90 ways of setting the RQA has been tested.

The most time demanding operation in this approach was the search of threshold belonging to considered RR, while the 10 approaches of defining lmin were computed very fast afterwards. The search of threshold for some RR can be programmed in many ways depending on many factors and this technical point can be the point of future research. In this study the way of searching was starting every time from threshold equal zero with the step derived from the variance and when reaching the region close to desired RR, the step was divided in loop by 10 in order not to jump over the desired precision. The total time taken for the entire computation/calculations estimated $\sim$ 100 hours, including the process of saving of the intermediate results, using the available computation facilities.

The measure of the ability to sorting the ALCs  according to the deterministic content (estimated by mDET measure) has been defined in two ways:
 a) by putting more significance to the ordering of stronger signals  - when the absolute value of the difference of computed order by a RQA setup and the  order of the 10 values of SNR ratios [0.005, 0.01, 0.025, 0.5, 0.75, 1 ,1.5, 2, 3, 5]  is summed,  b) the difference between the places in ordering of  10 ALCs by a RQA setup and the vector of the defined positions [1, 2, 3, 4, 5, 6, 7, 8, 9, 10] (ascending in our code)  in absolute value is summed, by putting less significance between the signals strengths as in this case the stronger signals would not contribute more to the measure. 
 The two  measures, when taking into account 10 ALCs with different SNR, can mathematically be expressed as:
 \beq
 measure  =  \sum_{i = 1}^{10} | x_i - y_i|,
 \eeq
 where $x_i$ is the i-th element of  the vector $x$  denoting  the SNR/position of ALC according it's  mDET measure  computed by some setup of RQA, and $y_i$ is the element of the vector of  $y$ of  the SNRs/positions as  defined. Most naturally, the  $ x_i - y_i$ are subtracted either when vector  $y$  is ordered  in descending or ascending fashion.
 
 One can observe how this measures works in Table \ref{TAB4}, which shows that in this particular case, the measure1 = $ \sum |[0.005 - 0.010, 0.01 - 0.005, 0.025 - 0.025, 0.5 - 1, 0.75 - 1.5, 1 - 2 ,1.5 -  0.75, 2 - 0.5, 3 - 5 , 5 - 3]|    $ = 8.51 and  measure2 =   $ \sum |[1 - 2 , 2 - 1, 3 - 3, 4 - 6, 5 - 7, 6 - 8, 7 - 5, 8 - 4, 9 - 10 , 10 - 9]|    $ = 16.   Consequently, we consider the 10 particular SNRs, the measures for which are bounded in the intervals, where measure1 $ \in [0,22.42] $  and measure2 $ \in [0,50]$\label{measures}. The important results from above described testing are presented in the Table \ref{TAB3}, where columns of measure2 and  measure1 averaged for 5 different generators of ALCs. The best performances for considered RRs are provided in the table, it can be observed that the second table on the right has measures1 and 2 higher which suggests that the ability to recognize between the different SNR ratios slightly worsens in case of interpolated data .

The lmin  adjusted to RR performs better than fixed ones especially when averaged to lower RRs $\leq 5 \%$ for both data sets.  Naturally in this case the lmin adjusts itself only with a few RRs. 
 For interpolated values the adjusted lmin works better for most of the RRs.
For data with gaps the fixed lmin shows better performance above RR =  5 \% and for both data sets better performance when averaged to the highest values of RR = 95 \%. 

When differing between the performance on the ALCs generated by the same setup for 10 different  SNRs and the ALCs, where every from 10 ALCs has different randomized generator the ability to sort by SNR is naturally worse for the randomized ones, and interestingly the methods with adjusted lmin perform better than the fixed ones. The randomized generators of ALCs are in the training sets represented by 2/5 with the assumption that the blazar variability might be governed by similar underlying  processes. The Table \ref{TAB3} shows the results applied on all 5 ALCs generators. 

The configuration with lowest measure1 and 2 for data with gaps appears for RR = 2 \%, where
   lmin\# = 7, emb\# = 3, tau\# = 1,  denoting the adjusted lmin scales with RR by the rule lmin = RR + 2, 
with $\tau$ and $m$ taken as maximum from the calculated time lags and embeddings and this configuration applied on real data is presented in Table \ref{TAB1}.

The configuration with lowest measure1 and 2 for data with interpolated values appears for RR = 3 \%, where   lmin\# = 7, emb\# = 3, tau\# = 1,  so the lmin scales the same as with data with gaps, 
with $\tau$ value as its own for every Lc and  $m$ taken again as maximum from the calculated embeddings and this configuration applied on real data is presented in Table \ref{TAB2}.

The best  configuration  with fixed lmin averaged till RR = 95 \% for  data with gaps appears for   lmin\# = 2, emb\# = 3, tau\# = 1,  where the lmnin = 3, with $\tau$ and $m$ chosen as in previous case and this configuration applied on real data is presented in Table \ref{TAB5}.


\begin{table*}[ht]
\centering
\begin{tabular}{l|rrrrr|rrrrr}
 \hline
    RR&  & & Part A &  & {\color{white}Part A} &  & &Part B  & \\ 
 \%& lmin$\#$ & emb$\#$ & tau$\#$ & measure1 & measure2 & lmin$\# $& emb$\# $& tau$\#$ &measure1 & measure2 \\ 
  \hline
1 & 6 & 2 & 3 & 12.43 & 26.67 & 9 & 3 & 1 & 13.57 & 28.00 \\ 
 2 & \textbf{7} & \textbf{3} & \textbf{3} & \textbf{10.35} & \textbf{25.00} & 1 & 3 & 1 & 11.35 & 27.23 \\ 
 3 & 7 & 3 & 3 & 11.72 & 26.33 & \textbf{7} & \textbf{3} &\textbf{ 1} &\textbf{ 11.12} &\textbf{ 26.12} \\
 4 & 7 & 3 & 3 & 11.72 & 27.00 & 7 & 3 & 1 & 12.78 & 26.33 \\ 
 5 & 6 & 2 & 2 & 13.99 & 29.33 & 10 & 3 & 1 & 14.12 & 29.33 \\ 
 10 & 2 & 3 & 3 & 13.12 & 31.33 & 4 & 3 & 2 & 14.93 & 34.67 \\ 
 15 & 1 & 2 & 3 & 12.48 & 28.00 & 4 & 3 & 2 & 14.93 & 33.33 \\ 
 20 & 1 & 3 & 3 & 12.43 & 30.67 & 3 & 3 & 1 & 14.77 & 32.67 \\ 
 25 & 1 & 1 & 3 & 11.98 & 28.00 & 5 & 2 & 1 & 14.93 & 33.33 \\ 
  30 & 1 & 1 & 3 & 11.97 & 27.33 & 5 & 2 & 1 & 14.93 & 33.33 \\ 
 35 & 2 & 3 & 3 & 12.27 & 30.67 & 9 & 3 & 1 & 14.77 & 33.33 \\ 
 40 & 2 & 2 & 3 & 12.27 & 28.00 & 6 & 3 & 1 & 14.61 & 33.33 \\ 
45 & 2 & 2 & 1 & 12.27 & 28.00 & 9 & 1 & 1 & 14.76 & 32.67 \\ 
 50 & 2 & 1 & 1 & 12.27 & 28.00 & 9 & 2 & 3 & 14.76 & 32.67 \\ 
 55 & 2 & 1 & 1 & 12.27 & 28.00 & 3 & 3 & 3 & 14.11 & 33.33 \\ 
 60 & 2 & 1 & 1 & 12.27 & 28.00 & 10 & 3 & 3 & 14.11 & 34.00 \\ 
 65 & 2 & 3 & 1 & 12.61 & 29.33 & 10 & 3 & 3 & 14.11 & 34.00 \\ 
 70 & 2 & 3 & 1 & 12.27 & 28.67 & 10 & 3 & 1 & 14.11 & 34.00 \\ 
 75 & 2 & 3 & 1 & 12.27 & 28.67 & 10 & 3 & 1 & 14.11 & 34.00 \\ 
 80 & 2 & 3 & 1 & 12.27 & 28.67 & 9 & 3 & 1 & 14.35 & 34.67 \\ 
85 & 2 & 3 & 1 & 12.27 & 28.67 & 7 & 3 & 1 & 14.35 & 35.33 \\ 
 90 & 2 & 3 & 1 & 12.61 & 29.33 & 3 & 3 & 1 & 14.94 & 34.67 \\ 
95 & 2 & 3 & 1  & 12.27 & 28.67 & 3 & 3 & 1 & 14.44 & 36.00 \\ 
   \hline
\end{tabular}
\caption{The table showing the best performances of 90 tested configurations of averaged RQA for different highest averaged RR [\%] listed in first column. 
 Part A of the table corresponds to the artificial light curves  with introduced gaps, and  Part B belongs the results of the data with linearly interpolated gaps. 
The lmin\#  column takes integer numbers from 1-10 and denotes setup of minimal line length, where 
lmin\#  1-4 belongs to fixed lmin of values equal to 2, 3, 4, 5 and lmin\# 5-10 belongs to lmin adjusted with RR.    Lmin6 is scaled according to the RR by the rule lmin5 = RR[\%], lmin6 = RR[\%] + 1, the lmin7 = RR[\%] + 2 and similarly up to lmin10.  
 The columns emb\# and tau\#  have been considered for testing  of the set of ALCs as 1 -  individual values corresponding to  each ALC, 2 - the mean of the whole considered set, 3 - the maximal value from the considered set of 10  ALCs with SNR ratio of values from [0.005, 0.01, 0.025, 0.5, 0.75, 1 ,1.5, 2, 3, 5].
The column measure1 and measure2  correspond to the summed absolute value of difference between vector of method ordered set of 10 ALCs and the real defined ascending order (the measures are described in Section \ref{measures}). The values in bold correspond to best - lowest scoring configurations and its measures, while measure1 $ \in [0,22.42] $  and measure2 $ \in [0,50]$.}
 \label{TAB3}
\end{table*}

\begin{table*}[ht]
\centering
\begin{tabular}{lrrrrrrrrrr}
  \hline
 & SNR & Len & $\tau$ & m & mD & mL & mEN & msD & msL & msEN \\ 
  \hline
1 & 3.000 & 352 & 3 & 8 & 0.789 & 52.25 & 0.12 & 10.00 & 1.66 & 1.00 \\ 
  2 & 5.000 & 446 & 4 & 5 & 0.785 & 86.24 & 0.11 & 9.00 & 2.66 & 2.00 \\ 
  3 & 0.500 & 265 & 5 & 8 & 0.581 & 105.21 & 0.05 & 8.00 & 3.66 & 3.00 \\ 
  4 & 0.750 & 302 & 4 & 6 & 0.530 & 142.22 & 0.15 & 7.00 & 4.66 & 4.00 \\ 
  5 & 2.000 & 332 & 5 & 9 & 0.414 & 122.14 & 0.14 & 5.66 & 5.66 & 5.00 \\ 
  6 & 1.500 & 375 & 6 & 4 & 0.412 & 110.87 & 0.15 & 4.66 & 3.00 & 9.00 \\ 
  7 & 1.000 & 362 & 3 & 5 & 0.358 & 75.66 & 0.08 & 4.33 & 6.66 & 6.00 \\ 
  8 & 0.025 & 380 & 3 & 7 & 0.342 & 120.45 & 0.09 & 3.33 & 8.00 & 7.33 \\ 
  9 & 0.005 & 402 & 5 & 7 & 0.152 & 140.54 & 0.05 & 2.00 & 9.00 & 8.33 \\ 
  10 & 0.010 & 245 & 2 & 10 & 0.144 & 88.14 & 0.07 & 1.00 & 10.00 & 9.33 \\ 
   \hline
   \hline
\end{tabular}
\caption{This table is example from the testing/parameter search process for the RQA input for the real data. Showing the configuration of  lmin\# = 7, emb\# = 3, tau\# = 3, where RQA measures are averaged up to 2 \% of RR, which came as the most suitable value based on the measure1 and 2 in Table \ref{TAB3} for data with gaps, applied to 10 artificial light curves generated by the same configuration, which differs in signal to noise ratio denoted as SNR and the lengths.   Out of the 513 number of generated data points,  up to 50 \% were deleted in order to mimic the real data sampling. The table is ordered according to the decreasing order of the mD (mean determinism) column.
}
 \label{TAB4}
\end{table*}

\begin{figure*}
\centering
\includegraphics[width=170mm,scale=1]{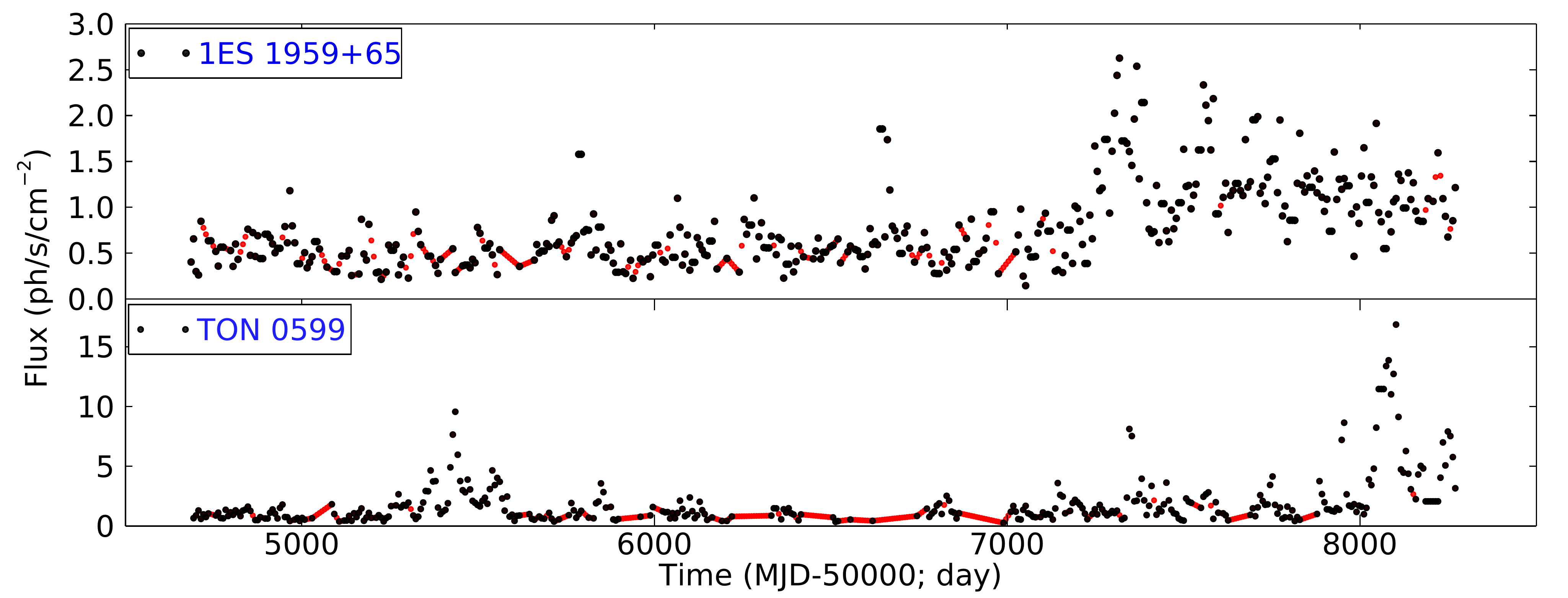}
\caption{The weekly binned Fermi/LAT observations (black symbols) were made evenly spaced by  the linear interpolation (red symbols). The upper and lower panels show the light curve of the blazars 1ES 1959+65 and TON 059, respectively. The uncertainty in the flux are not shown here for the clarity.}
\label{ILC}
\end{figure*}

\begin{table*}[ht]
\centering
\begin{tabular}{llrrrrrrrrr}
  \hline
 & source & Len & $\tau$ & m & mD & mL & mEN & msD & msL & msEN \\ 
  \hline
1 & PKS 1502+106 & 384 & 6 & 9 & 0.8649046 & 11.5815317 & 1.7666925 & 18.05 & 8.60 & 8.40 \\ 
  2 & CTA 102 & 425 & 6 & 10 & 0.8142809 & 16.5577447 & 1.6067981 & 16.95 & 6.95 & 6.60 \\ 
  3 & 4C+38.41 & 462 & 7 & 11 & 0.7975748 & 9.9117767 & 1.7781942 & 14.65 & 7.40 & 9.25 \\ 
  4 & PKS 1424-418 & 473 & 7 & 10 & 0.7774705 & 13.8187285 & 1.9607682 & 15.20 & 8.40 & 10.05 \\ 
  5 & AO 0235+164 & 273 & 4 & 11 & 0.7586913 & 14.8260550 & 1.4298484 & 14.40 & 8.30 & 5.10 \\ 
  6 & 3C 454.3 & 462 & 9 & 9 & 0.7497014 & 10.3332515 & 1.7983391 & 15.20 & 5.80 & 9.20 \\ 
  7 & Mrk 421 & 509 & 6 & 8 & 0.7406063 & 33.5535961 & 1.6550572 & 14.30 & 8.20 & 7.40 \\ 
  8 & TON 0599 & 355 & 7 & 13 & 0.7332010 & 18.2211392 & 1.7143814 & 13.50 & 6.50 & 6.25 \\ 
  9 & 3C 279 & 502 & 8 & 8 & 0.7042435 & 11.5690254 & 1.7866271 & 12.05 & 6.55 & 9.00 \\ 
  10 & 1ES 1959+65 & 420 & 5 & 8 & 0.6961874 & 28.9787136 & 1.4342795 & 10.70 & 7.80 & 5.00 \\ 
  11 & 4C+21.35 & 373 & 3 & 7 & 0.6824060 & 69.2425815 & 4.0358228 & 9.65 & 19.25 & 18.90 \\ 
  12 & Mrk 501 & 461 & 2 & 9 & 0.6432987 & 34.0905747 & 3.7094471 & 8.75 & 15.00 & 17.20 \\ 
  13 & PKS 0454-234 & 472 & 3 & 10 & 0.5992828 & 46.7527960 & 3.8563625 & 7.10 & 17.30 & 17.45 \\ 
  14 & PKS 2155-304 & 507 & 6 & 9 & 0.5970160 & 7.7267807 & 1.2505630 & 8.70 & 2.35 & 3.20 \\ 
  15 & S5 0716+714 & 490 & 4 & 9 & 0.5767199 & 6.7950413 & 0.6800733 & 7.45 & 3.10 & 2.25 \\ 
  16 & 3C 273 & 363 & 3 & 8 & 0.5619126 & 74.9299485 & 3.1556767 & 7.40 & 16.90 & 13.50 \\ 
  17 & W Comae & 208 & 2 & 6 & 0.5556650 & 35.9415403 & 3.1834867 & 5.50 & 13.65 & 13.35 \\ 
  18 & 3C 66A & 494 & 3 & 9 & 0.5444408 & 51.3307591 & 3.7218713 & 4.70 & 17.20 & 17.00 \\ 
  19 & BL Lac & 475 & 3 & 9 & 0.5164275 & 43.1954701 & 3.8138898 & 4.50 & 16.55 & 17.70 \\ 
  20 & ON+325 & 447 & 3 & 8 & 0.3585777 & 51.9465621 & 3.1629330 & 1.25 & 14.20 & 13.20 \\ 
   \hline
\end{tabular}
\caption{The non-linear time series analysis applied on the \gama-ray light curves of 20 blazars as in Table \ref{TAB1}, but here the mean averaged RQA measures are averaged till RR = 95 \% with the step of 5 \% and the embedding  $m $ taken as maximal value from the set of calculated embeddings and the $\tau $  taken as own value for every observation.
The distinction between FSRQs and BL Lacs is also significant. In comparison with table  \ref{TAB1} the mean DET (mD) and mean ENTR (mEN) gained some value as it is averaged to high values of RR.
}
 \label{TAB5}
\end{table*}



\end{document}